\documentclass{article} 
\usepackage{graphicx}
\graphicspath{{converted_graphics/}}
\begin{document}

\begin{center}

{\LARGE Observations of an Impurity-driven Hysteresis}\vskip6pt 
{\LARGE Behavior in Ice Crystal Growth at Low Pressure}\vskip16pt

{\Large Kenneth G. Libbrecht}\footnote{%
e-mail address: kgl@caltech.edu\newline
}\vskip4pt

{\large Department of Physics, California Institute of Technology}

{\large Pasadena, California 91125}

\vskip18pt

\hrule \vskip1pt \hrule
\vskip 14pt
\end{center}

\noindent \textbf{Abstract. We describe observations of a novel hysteresis
behavior in the growth of ice crystals under near-vacuum conditions. Above a
threshold supersaturation, we find that the ice growth rate often exhibits a
sudden increase that we attribute to an impurity-driven growth instability.
We examine possible mechanisms for this instability, which can be used to
produce clean, faceted ice surfaces. }

\section{\noindent Introduction}

The formation of complex structures during solidification often results from
a subtle interplay of nonequilibrium, nonlinear processes, for which
seemingly small changes in molecular dynamics at the nanoscale can produce
large morphological changes at all scales. One popular example of this
phenomenon is the formation of snow crystals, which are ice crystals that
grow from water vapor in an inert background gas. Although this is a
relatively simple physical system, snow crystals display a remarkable
variety of columnar and plate-like forms, and much of the phenomenology of
their growth remains poorly understood, even at a qualitative level \cite%
{libbrechtreview}.

\begin{figure}[tbp] 
  \centering
  \includegraphics[bb=0 0 2705 2720,width=4in,keepaspectratio]{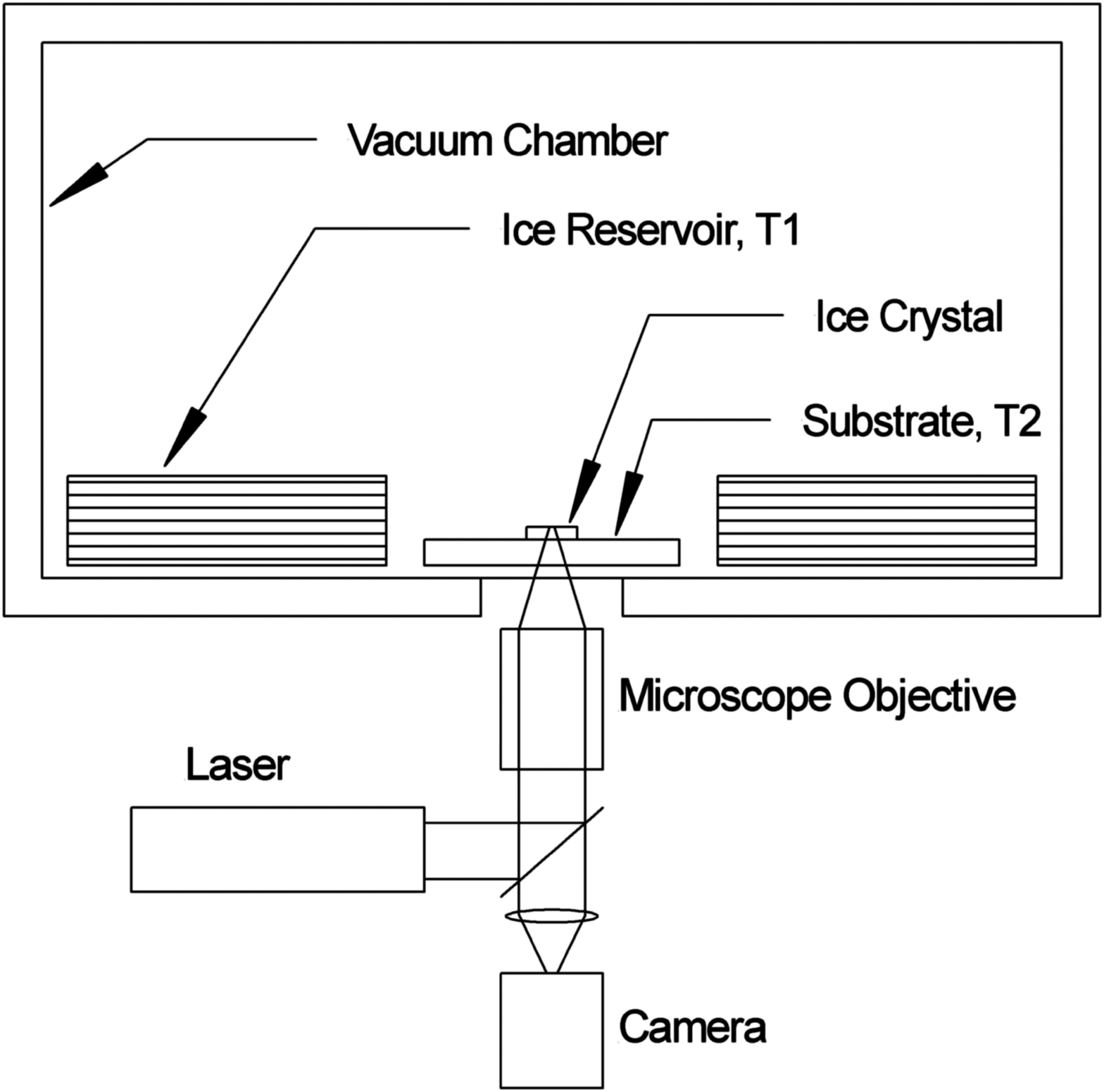}
  \caption{The basic layout of our
experimental apparatus. An ice crystal is placed with known orientation on a
substrate inside an evacuated growth chamber. An ice reservoir inside the
chamber provides a source of water vapor to grow the sample crystal. The
supersaturation is determined by the temperature difference between the ice
reservoir (equal to the temperature of the rest of the growth chamber) and
the substrate. The sample crystal is imaged using a microscope objective and
a video camera. A low-power laser is focused onto the crystal by the same
microscope objective. The laser spot is reflected by the top and bottom
surfaces of the ice crystal, and the brightness of the two interfering beams
oscillates as a function of crystal thickness \cite{precision}. }
  \label{basicschematic}
\end{figure}

As we described in two earlier papers \cite{principal, precision}, we have
been investigating the growth behavior of ice under near-vacuum conditions
in order to better understand the surface attachment kinetics that governs
the crystal formation process. Toward this end, we have constructed an
experimental apparatus that allows us to place single, faceted ice crystals,
typically 10-50 $\mu $m in size, on a temperature-controlled substrate
surrounded by an ice reservoir in a small vacuum chamber, as shown in 
Figure \ref{basicschematic} \cite{precision}.

We write the growth velocity $v_{n}$ normal to the surface of a crystal
facet in terms of the Hertz-Knudsen formula 
\begin{equation}
v_{n}=\alpha v_{kin}\sigma   \nonumber
\end{equation}%
where $v_{kin}$ is a temperature-dependent kinetic velocity derived from
statistical mechanics, $\sigma \ $is the water vapor supersaturation just
above the growing surface, and $\alpha $ is known as the condensation
coefficient, which contains the attachment kinetics that governs how water
molecules are incorporated into the ice lattice \cite{libbrechtreview}.

The supersaturation in these experiments is given by%
\[
\sigma =\frac{c_{sat}(T_{reservoir})-c_{sat}(T_{substrate})}{%
c_{sat}(T_{substrate})} 
\]%
where $c_{sat}(T)$ is the saturated vapor pressure of ice at temperature $T$%
, while $T_{reservoir}$ and $T_{substrate}$ are the temperatures of the ice
reservoir and crystal sample, respectively (see Figure \ref{basicschematic}).

We use laser interferometry to measure the growth rate $v_{n}$, from which
we can then extract $\alpha $ as a function of $\sigma $, $T_{substrate}$,
and other experimental parameters. The measurements shown here were of the
basal surfaces of thin, plate-like crystals growing near $T_{substrate}=-15$
C with a background air pressure near 5 mbar. At this pressure and with the
small ice crystals observed, the crystal growth is not limited by molecular
diffusion through the surrounding air, but is rather limited by attachment
kinetics on the ice surface. Additional details describing our apparatus and
important systematic errors can be found in \cite{precision}.

\begin{figure}[tbp] 
  \centering
  \includegraphics[width=4.5in,height=5.44in,keepaspectratio]{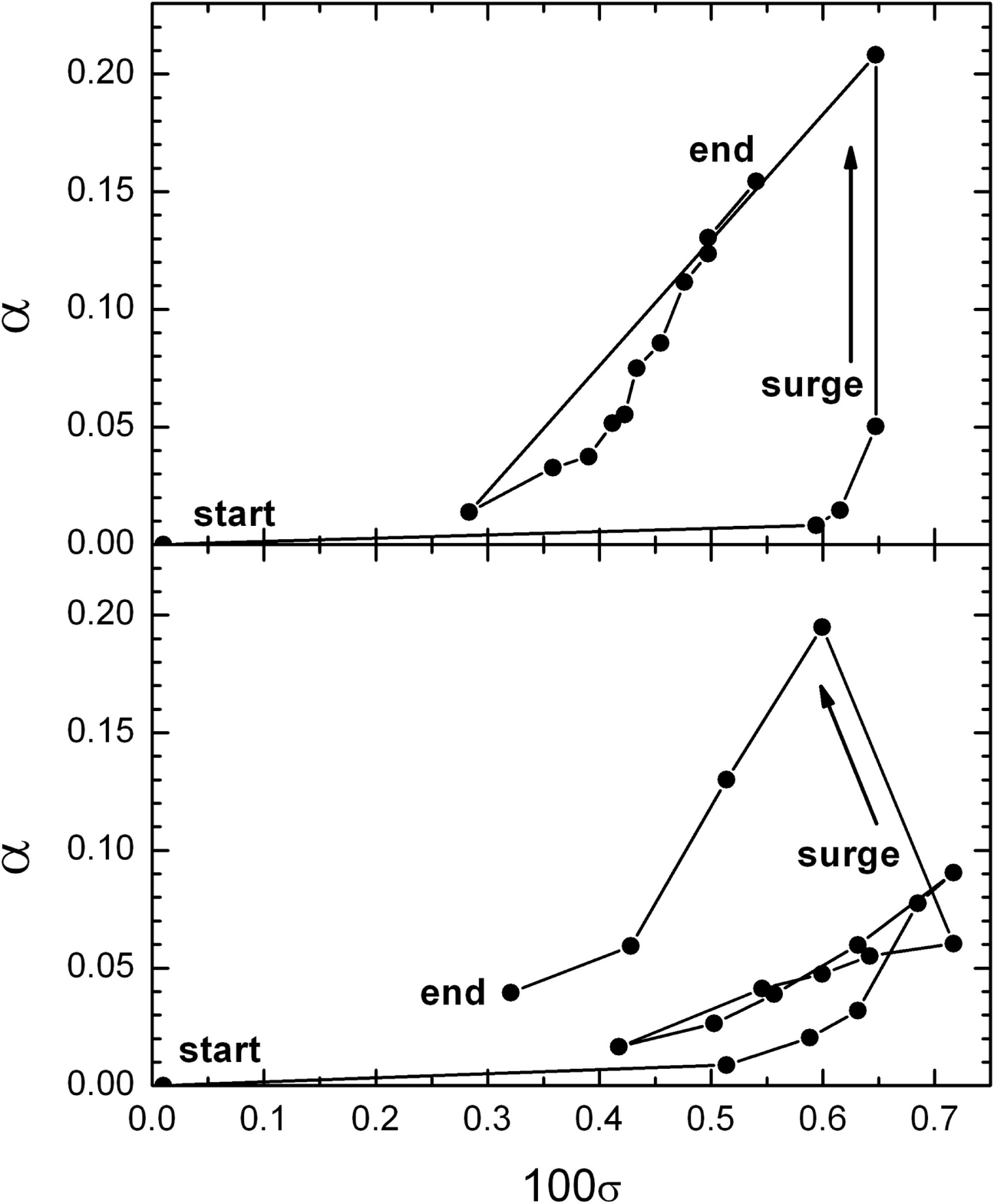}
  \caption{Two examples of the
impurity-driven growth instability described in the text. Both panels show a
time series of measurements of the condensation coefficient $\protect\alpha $
for the basal facet a single plate-like ice crystal as a function of
supersaturation $\protect\sigma $. The background pressure in the
experiments was approximately 5 mbar and the crystal temperature was near
-15 C. In both cases the growth was relatively slow until the occurrence of
a sudden \textquotedblleft surge\textquotedblright\ of growth, after which $%
\protect\alpha \left( \protect\sigma \right) $ remained substantially higher
than it was before the surge.}
  \label{examples}
\end{figure}

\begin{figure}[tbp] 
  \centering
  \includegraphics[width=4.5in,height=3.48in,keepaspectratio]{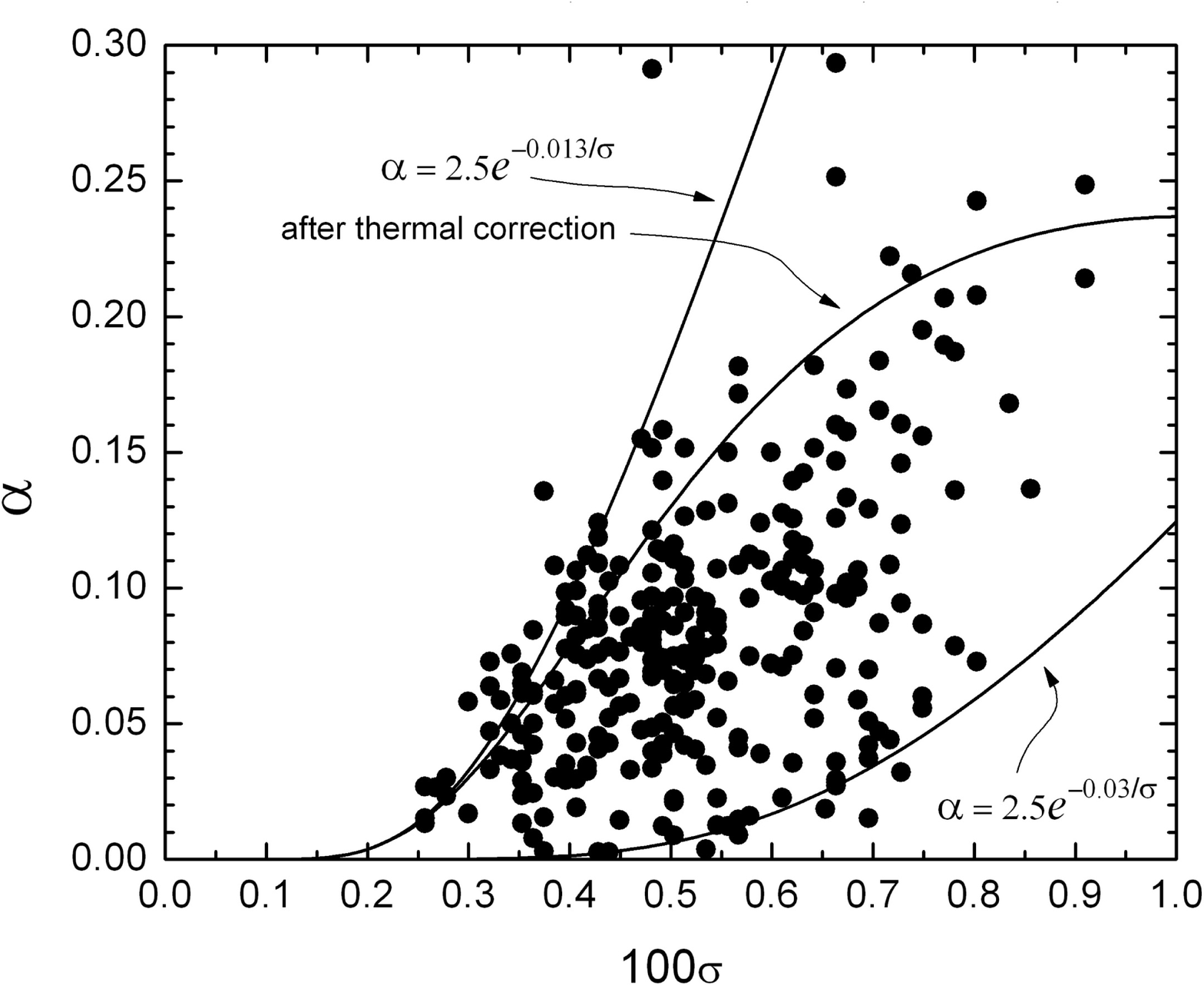}
  \caption{Combined measurements
of eleven crystals, which includes all the crystals from a single set of
observations. The higher points in this figure were all from
\textquotedblleft post-surge\textquotedblright\ measurements like those
shown in Figure 2. The lower curve shows an approximate lower bound for the
measurements, while the upper curve gives an approximate upper bound for
post-surge measurements. The middle curve gives the upper bound after
correcting for surface heating as described in \protect\cite{libbrechtreview}%
. The data show a fairly distinct upper bound in the growth, which we
interpret as the intrinsic growth rate of an impurity-free basal surface at
-15 C.}
  \label{compplot}
\end{figure}

\section{Observations}

In the course of our experiments we observed an unexpected hysteresis
behavior, of which we show two examples in Figure \ref{examples}. In both
cases shown, an ice crystal was first placed on the substrate while the
chamber pressure was near one bar, as described in \cite{precision}. The
chamber pressure was then slowly reduced while keeping $\Delta
T=T_{reservoir}-T_{substrate}\approx 0$ to prevent significant growth or
evaporation of the crystal. This process typically took 3-5 minutes, ending
with the chamber pressure near 5 mbar and the crystal at the
\textquotedblleft start\textquotedblright\ location shown in Figure \ref%
{examples}.

Once the pressure and temperatures were stable, we then slowly reduced $%
T_{substrate}$ to increase $\sigma $ while observing the crystal growth
velocity. Once we obtained measurable growth, we adjusted $T_{substrate}$
(and thus $\sigma $) as a function of time and observed the growth behaviors
shown in Figure \ref{examples}. The data-taking process took several minutes
and usually ended when the crystal was deemed to large to continue \cite%
{precision}.

We observed considerable crystal-to-crystal variability in our data, but the
following features shown in Figure \ref{examples} were often seen:

1) Crystal growth did not commence at a measurable rate until $\sigma $ was
fairly high, above roughly 0.5\% for the examples shown.

2) The measured $\alpha (\sigma )$ remained relatively low for the early
stages of growth. This is best seen in the second example in Figure \ref%
{examples}, where $\sigma $ was varied up and down while observing the
crystal growth.

3) When $\sigma $ was raised sufficiently, the growth exhibited a sudden
\textquotedblleft surge\textquotedblright , as labeled in the Figure. The
surge occurred in just a few seconds and increased the growth rate to nearly
1 $\mu $m/sec in some cases, implying $\alpha \approx 1$ at the peak growth.
Calculations show that such rapid growth produced local heating at the
crystal surface \cite{libbrechtreview} that also limited the growth. Upon
seeing the growth surge, $\sigma $ was typically reduced to keep the crystal
from growing too large.

4) After the spike in growth, the growth profile $\alpha (\sigma )$ remained
substantially higher than the pre-surge profile.

5) If the supersaturation was reduced to $\sigma =0$ for several minutes, or
if a small quantity of gas was let into the chamber and pumped out, then $%
\alpha (\sigma )$ returned to lower values.

Not all crystals exhibited a growth surge, while some did so clearly. Figure %
\ref{compplot} shows a set of all measurements taken during a single
experimental run measuring basal growth at -15 C. The data show a rough
lower bound, in that all crystals started growing once the supersaturation
approached one percent. The data also show a rather distinct upper bound
profile $\alpha (\sigma )$ that was not exceeded by any post-surge data.

\section{Interpretation}

We interpret these observations with a simple qualitative model involving
impurities on the ice surface. After being nucleated in air at one bar \cite%
{precision}, a typical sample crystal grows for 1-2 minutes before being
placed in the vacuum chamber, and then it sits on the substrate for 3-5
minutes during the pump-out period. We believe that impurities build up on
the ice surface during this time, which reduces $\alpha (\sigma )$. Because
of this, $\sigma $ must be raised quite high to produce the first measurable
growth.

The impurity surface density is then apparently reduced as ice is deposited
during growth (we discuss possible mechanisms for this below). For low
growth rates, this surface cleaning is not very efficient, so $\alpha
(\sigma )$ remains low, and the surface impurities mostly remain on the
surface as they are rejected from being incorporated into the ice lattice.
At a sufficiently high $\sigma $, however, the impurity cleaning process
leads to a runaway growth instability. As impurities are removed from the
surface, the growth rate increases on the now cleaner surface, which thus
accelerates the removal of impurities. The resulting positive feedback
causes the growth rate to suddenly increase, producing a growth surge. Once
the surface has been cleaned by a period of rapid crystal growth, it remains
relatively clean in the low-pressure environment, so $\alpha (\sigma )$
remains high. If the growth is halted for some period of time, however, then
gaseous impurities will deposit and build up on the surface, decreasing $%
\alpha (\sigma )$ once again.

\section{Impurity Removal Mechanisms}

Assuming this overall picture is correct, there are several mechanisms for
which crystal growth can remove surface impurities:

\textbf{1) Incorporation.} As ice is deposited during growth, adsorbed
impurity molecules may become buried in the bulk. This would remove
impurities from the surface, and we expect this process would increase with
increasing growth rate. It is well known, however, that impurities are not
readily incorporated into the lattice when ice forms from liquid water, so
we expect that foreign molecules are not easily buried during ice growth
from water vapor either. Unfortunately, our understanding of this process is
not sufficient to allow us to make quantitative estimates for impurity
incorporation rates.

\textbf{2) Desorption.} It is also possible that faster growth can cause
impurities to become desorbed from the surface. In one sense this seems
unlikely; ice has a high vapor pressure, which means that water molecules
are continually condensing onto and evaporating off of the ice surface.
During our measurements, these two processes are out of equilibrium by less
than one percent (equal to $\sigma $), and it seems unlikely that such a
small imbalance in growth and evaporation would lead to impurity desorption.
On the other hand, growth will tend to push foreign molecules up to keep
them from being incorporated into the lattice. This process could weaken the
surface binding, and at sufficiently high growth rates it could cause
desorption. Thus we cannot rule out this mechanism for impurity removal.

\textbf{3) Flow.} If impurity molecules are sufficiently mobile on the
surface, they may also be swept laterally by expanding molecular terraces as
the crystal grows. This process would push foreign molecules to the edges of
small facets, and for the observed basal growth it would deposit impurities
on the prism surfaces. In several cases we have seen small plate-like
crystals grow upward into fairly tall columns (aspect ratios > 5) 
as the rapid basal growth was accompanied by little growth of the prism
surfaces. This odd behavior was unexpected at -15 C, where thin plates
normally grow in air, but it would be explained by the impurity flow
mechanism.

\textbf{4) Combinations.} The above mechanisms could also work in
combination. For example, foreign molecules could flow to regions of high
impurity density, where they could desorb or become buried. We have commonly
seen crystals with slow-growing regions (for example, see Figure 2 in \cite%
{precision}) that appear to have a high concentration of surface impurities.
It may be that surface flow is depositing impurities in these regions.

Whatever the mechanism, we see from Figure \ref{examples} that growth
velocities of order $v_{n}\approx 0.1$ $\mu $m/sec are sufficient to engage
this instability, at least under the conditions of our measurements.

\section{Discussion}

Our growth data showed considerable crystal-to-crystal variability, and the
growth \textquotedblleft surge\textquotedblright\ with its accompanying
hysteresis was not always present. However, in many cases the surge was
clear and unmistakable when seen in real time. We believe that the
impurity-driven instability described above explains the data
satisfactorily, and differences in impurity levels could easily explain the
crystal-to-crystal variability as well.

These observations also suggest that our previous data were contaminated by
impurities to some degree. In \cite{principal}, our experiment was new and
our impurity problems probably the most substantial. We rebuilt our
apparatus for \cite{precision} and put additional effort into producing a
cleaner growth chamber. The current experiment uses the same hardware as 
\cite{precision}, but additional time has allowed some further outgassing of
impurities. A comparison of these experiments indicates that $\alpha $ has
steadily increased as the hardware has become cleaner.

Throughout these measurements we typically parameterized our data with $%
\alpha \rightarrow A\exp (-\sigma _{0}/\sigma )$, and in doing so we found
that most of the variability was in $A.$ The critical supersaturation $%
\sigma _{0}$ does not seem to change greatly with the addition of surface
impurities. This implies that surface impurities do not substantially change
the edge free energy $\beta $ for a molecular terrace or the 2D nucleation
process in general \cite{principal}. Instead, it appears that surface
impurities reduce the growth of 2D islands, which then reduces $A.$ We will
discuss mechanisms for this in a subsequent paper.

It remains something of a puzzle why the data in \cite{precision} were so
self-consistent, since subsequent data (for example Figure \ref{compplot})
have shown considerably higher crystal-to-crystal variability. The
measurements in \cite{precision} were taken soon after our new chamber was
constructed, so we speculate that the walls were still outgassing
sufficiently to reduce $\alpha $ for the crystals we measured. We also
suspect that the higher impurity levels suppressed growth surges. We are
planning additional experiments in which impurities are systematically added
to our growth chamber in order to investigate these effects further.

The overarching conclusions we draw from these observations are that even
relatively low levels of surface impurities may have a substantial impact on
ice crystal growth, and that the presence of surface impurities generally
seems to reduce ice growth rates.

\end{document}